\documentclass[12pt]{article}
\usepackage{graphicx}

\begin{document}

\title{Dynamics of light hypernuclei in collisions of $^{197}$Au+$^{197}$Au at GeV energies}

\author{Zhao-Qing Feng \footnote{Corresponding author. \newline \emph{E-mail address:} fengzhq@scut.edu.cn}}
\date{}
\maketitle

\begin{center}
\small \emph{School of Physics and Optoelectronics, South China University of Technology, Guangzhou 510640, China}
\end{center}

\textbf{Abstract}
\par
The dynamics of light hypernuclei and nuclear clusters produced in $^{197}$Au+$^{197}$Au collisions has been investigated thoroughly with a microscopic transport model. All possible channels of hyperon production and transportation of hyperons in nuclear medium are implemented into the model. The light complex fragments are recognized with the Wigner density approach at the stage of freeze out in nuclear collisions. The isospin diffusion in the collisions is responsible for the neutron-rich cluster formation. The collective flows of nuclear clusters are consistent with the experimental data from FOPI collaboration. It is found that the influence of the hyperon-nucleon potential on the free hyperons is negligible, but available for the light hypernuclide formation. The directed and elliptic flows of $^{3}_{\Lambda}$H and $^{4}_{\Lambda}$H at incident energies of 2, 2.5, 3, 3.5 and 4 GeV/nucleon are investigated thoroughly and manifest the same structure with the nuclear clusters. The hypernuclear yields are produced in a wide rapidity and momentum regime with increasing the beam energy.

\emph{PACS}: 21.80.+a, 24.10.Lx, 25.70.Pq     \\
\emph{Keywords:} Strangeness dynamics; Hypernucleus formation; Heavy-ion collisions; LQMD transport model

\bigskip

\section{Introduction}

Since the first observation of $\Lambda$-hypernuclide in nuclear multifragmentation reactions induced by cosmic rays in 1950s \cite{Da53}, a remarkable progress has been obtained in producing hypernuclides via different reaction mechanism, such as the high-energy lepton (photon, electron) induced reactions, collisions of the hadron ($\pi^{\pm}$, K$^{\pm}$, proton, antiproton etc) on a nucleus, and nuclear fragmentation reactions with high energy heavy-ion collisions. The hypernuclear spectroscopy, lifetime, giant monopole resonance, cluster structure and weak decay of hypernuclei were investigated in theories \cite{Na14,Ga16,Lv18,Hi18} and in experiments with the meson or electron beams at different laboratories in the world, such as the Japan's National Laboratory for High Energy Physics (KEK), Japan Proton Accelerator Research Complex (J-PARC), INFN's Double Annular $\phi$ Factory for Nice Experiments (DA$\Phi$NE), Thomas Jefferson National Accelerator Facility (JLab), and Mainz Microtron (MAMIC) accelerator \cite{Ha06,Bo12a,Fe15a}. High-energy heavy-ion collisions in the laboratories provide a unique way to produce the exotic hypernuclei, i.e., the multiple strangeness hypernuclei, extremely proton-rich/neutron-rich hypernuclei, neutral few-body bound states such as nn$\Lambda$ \cite{Ra13a}. Recently, the antihypernuclide $^{3}_{\overline{\Lambda}}\overline{H}$ was found by the STAR collaboration in relativistic heavy-ion collisions \cite{Star}. The first successful experiment to produce hypernuclei with heavy-ion beams impinged on the fixed target was the HypHI experiment at GSI \cite{Ra13}.

The strangeness physics is also attracted attention in the future experiments at the high-intensity heavy-ion accelerator facility (HIAF) in China \cite{Ya13,Ch19}. With the establishment of new facilities in the world, e.g., PANDA (Antiproton Annihilation at Darmstadt, Germany) \cite{Pand} and Super-FRS/NUSTAR \cite{Sa12} at FAIR (GSI, Germany), NICA (Dubna, Russia) \cite{Nica}, J-PARC (Japan) \cite{Ta12}, the hypernuclear physics will be thoroughly investigated.
The proton-rich and neutron-rich hypernuclei in heavy-ion collisions is planned by the HypHI collaboration at the future facility for antiproton and ion research (FAIR) with super-FRS (FRagment Separator) \cite{Ra16}. A complex hypernuclear detector WASA-FRS is managed for measuring hyperfragments in heavy-ion collisions at GSI \cite{Sa16}.
On the other hand, the properties of high-baryon matter may be investigated via the strange particles in high-energy heavy-ion collisions for extracting the in-medium hadrons and nuclear equation of state (EOS) \cite{Ca99, Fu06, Li08, To10, Ha12}. Inclusion of the strangeness degree of freedom in nuclear matter and in a nucleus extends the research activities, in particular on the issues of the nuclear structure of hypernucleus and kaonic nucleus, hyperon-nucleon and hyperon-hyperon interactions, probing the in-medium properties of hadrons \cite{Gi95}. Recently, the hypertriton lifetime puzzle is investigated by using the three-body wave functions generated in a chiral effective field theory approach \cite{Pe20}. Moreover, hadrons with strangeness as essential ingredients influence the high-density nuclear EOS. The strangeness ingredient in dense matter softens the EOS at high-baryon densities, and consequently decreases the mass of neutron stars\cite{We12,Ji13}.

The formation of hypernuclei in heavy-ion collisions is associated the creation of hyperons in hadron-hadron collisions, transportation in nuclear medium, fragment recognition and statistical decay. Up to now, several models have been established for describing the hypernucleus production, i.e., the statistical multifragmentation model (SMM) \cite{Bo07,Bo12}, statistical approach with a thermal source \cite{An11} and microscopic transport models with different recognition methods for hypernucleus formation \cite{Bo15,Fe16,Gt16,Le19}. Some interesting results are obtained for describing the hypernucleus formation, i.e., the hyperfragment yields, excitation function of hyperfragment production, multiple strangeness hypernucleus etc. More sophisticated investigation of hypernuclear dynamics is needed in theories. The production of light hypernuclei in heavy-ion collisions manifests the 'fire ball' properties and nuclear structure. In this work, the Lanzhou quantum molecular dynamics (LQMD) transport model is extended for investigating the nuclear clusters and hypernuclear dynamics in heavy-ion collisions near threshold energies.

\section{The transport model and light fragment recognition}

In the LQMD transport model, the production of resonances with the mass below 2 GeV, hyperons ($\Lambda$, $\Sigma$, $\Xi$) and mesons ($\pi$, $\eta$, $K$, $\overline{K}$, $\rho$, $\omega$) is coupled in the reaction channels via meson-baryon and baryon-baryon collisions \cite{Fe11}. The model was used for investigating the density dependence of symmetry energy, isospin splitting of nucleon effective mass and in-medium effects of hadrons in heavy-ion collisions, spallation reactions induced by hadrons, annihilation mechanism in antiproton-nucleus collisions, see the review article in details \cite{Fe18}. Recently, the hypernuclear dynamics in hadron (antikaon, antiproton and proton) induced reactions \cite{Fe20}. The temporal evolutions of nucleons and nucleonic resonances are described by Hamilton's equations of motion under the self-consistently generated two-body and three-body interaction potential with the Skyrme-like force. The Hamiltonian of baryons consists of the relativistic energy, Coulomb interaction, momentum dependent potential and local interaction as follows:
\begin{equation}
H_{B}=\sum_{i}\sqrt{\textbf{p}_{i}^{2}+m_{i}^{2}}+U_{Coul}+U_{mom}+U_{loc}.
\end{equation}
Here the $\textbf{p}_{i}$ and $m_{i}$ represent the momentum and the mass of the baryons. The local interaction potential is evaluated from the energy-density functional of
\begin{equation}
U_{loc}=\int V_{loc}(\rho(\mathbf{r}))d\mathbf{r}
\end{equation}
 with
\begin{eqnarray}
V_{loc}(\rho)=&& \frac{\alpha}{2}\frac{\rho^{2}}{\rho_{0}} +
\frac{\beta}{1+\gamma}\frac{\rho^{1+\gamma}}{\rho_{0}^{\gamma}} + E_{sym}^{loc}(\rho)\rho\delta^{2}
\nonumber \\
&& + \frac{g_{sur}}{2\rho_{0}}(\nabla\rho)^{2} + \frac{g_{sur}^{iso}}{2\rho_{0}}[\nabla(\rho_{n}-\rho_{p})]^{2},
\end{eqnarray}
where the $\rho_{n}$, $\rho_{p}$ and $\rho=\rho_{n}+\rho_{p}$ are the neutron, proton and total densities, respectively, and the $\delta=(\rho_{n}-\rho_{p})/(\rho_{n}+\rho_{p})$ being the isospin asymmetry of baryonic matter. The parameters $\alpha$, $\beta$, $\gamma$, $g_{sur}$, $g_{sur}^{iso}$ and $\rho_{0}$ are set to be the values of -215.7 MeV, 142.4 MeV, 1.322, 23 MeV fm$^{2}$, -2.7 MeV fm$^{2}$ and 0.16 fm$^{-3}$, respectively. The set  of the parameters gives the compression modulus of K=230 MeV for isospin symmetric nuclear matter at the saturation density ($\rho_{0}=0.16$ fm$^{-3}$). The surface coefficients $g_{sur}$ and $g_{sur}^{iso}$ are taken to be 23 MeV fm$^{2}$ and -2.7 MeV fm$^{2}$, respectively. The third term contributes the symmetry energy being of the form $E_{sym}^{loc}=\frac{1}{2}C_{sym}(\rho/\rho_{0})^{\gamma_{s}}$. The parameter $C_{sym}$ is taken as the value of 38 MeV. The $\gamma_{s}$ could be adjusted to get the suitable case from constraining the isospin observables, e.g., the values of 0.5, 1 and 2 being the soft, linear and hard symmetry energy, respectively. Here, the linear symmetry energy is taken into account in the calculation. Combined the kinetic energy from the isospin difference of nucleonic Fermi motion, the three kinds cross at the saturation density with the value of 31.5 MeV. The Skyrme-type momentum dependent interaction is used in the hypernuclear description \cite{Fe11}.

The one-body potentials of kaons (antikaons) and hyperons are used for transportation in nuclear medium and evaluated by the chiral effective Lagrangian approach and relativistic mean-field theories \cite{Fe13,Fe15}, respectively. The optical potential of hyperon is written as
\begin{equation}
V_{Y}(\textbf{p},\rho)=\omega_{Y}(\textbf{p},\rho)-\sqrt{\textbf{p}^{2}+m^{2}},
\end{equation}
in which the in-medium energy is derived from
\begin{equation}
\omega_{Y}(\textbf{p}_{i},\rho_{i})=\sqrt{(m_{H}+\Sigma_{S}^{H})^{2}+\textbf{p}_{i}^{2}} + \Sigma_{V}^{H}.
\end{equation}
The self-energies of hyperons are assumed to be two thirds of that experienced by nucleons, namely $\Sigma_{S}^{\Lambda}= 2 \Sigma_{S}^{N}/3$, $\Sigma_{V}^{\Lambda}= 2\Sigma_{V}^{N}/3$, $\Sigma_{S}^{\Xi}= \Sigma_{S}^{N}/3$ and $\Sigma_{V}^{\Xi}= \Sigma_{V}^{N}/3$. The nucleon scalar $\Sigma_{S}^{N}$ and vector $\Sigma_{V}^{N}$ self-energies are computed from the well-known relativistic mean-field model with the NL3 parameter ($g_{\sigma N}$=8.99, $g_{\omega N}$=12.45 and $g_{\rho N}$=4.47) \cite{La97}.  The values of optical potentials at saturation density are -32 MeV and -16 MeV for $\Lambda$ and $\Xi$, respectively. A weakly repulsive $\Sigma-$N potential with 12 MeV at saturation density is used by fitting the calculations of chiral effective field theory \cite{Pe16}. The hyperon-nucleon interaction potentials will influence the dynamics of hyperons in nuclear medium, i.e., kinetic energy spectra, emission anisotropy etc. Furthermore, the bound states to form nuclear fragments and hypernuclei are modified by the potential.

Production and decay of the resonances below the mass of 2 GeV have been included in the model. All possible channels of $\Xi$ production in hadron-hadron collisions are implemented in this work. The strange particles are created in the direct process by the channels as follows
\begin{eqnarray}
&& BB \rightarrow BYK,  BB \rightarrow BBK\overline{K},  B\pi(\eta) \rightarrow YK,  YK \rightarrow B\pi,     \nonumber \\
&& B\pi \rightarrow NK\overline{K}, Y\pi \rightarrow B\overline{K}, \quad  B\overline{K} \rightarrow Y\pi, \quad YN \rightarrow \overline{K}NN,  \nonumber \\
&& BB \rightarrow B\Xi KK, \overline{K}B \leftrightarrow K\Xi, YY \leftrightarrow N\Xi, \overline{K}Y \leftrightarrow \pi\Xi
\end{eqnarray}
Here the symbols corresponding to B(N, $\triangle$, N$^{\ast}$), Y($\Lambda$, $\Sigma$), $\Xi(\Xi^{0}, \Xi^{-}$), $\pi(\pi^{-}, \pi^{0}, \pi^{+})$, K(K$^{0}$, K$^{+}$), $\overline{K}$($\overline{K}^{0}$, K$^{-}$). The elementary cross sections are parameterized by fitting the available experimental data and the Clebsch-Gordan coefficients for the isospin channels. Furthermore, the elastic scattering and strangeness-exchange reaction between strangeness and baryons have been considered through the channels of $KB \rightarrow KB$, $YB \rightarrow YB$ and $\overline{K}B \rightarrow \overline{K}B$ and we use the parametrizations in Ref. \cite{Cu90}. The charge-exchange reactions between the $KN \rightarrow KN$ and $YN \rightarrow YN$ channels are included by using the same cross sections with the elastic scattering, such as $K^{0}p\rightarrow K^{+}n$, $K^{+}n\rightarrow K^{0}p$ etc \cite{Fe13}.

For the fragment with $Z\leq$2, the Wigner phase-space density at freeze out is used to evaluate the probability of fragment formation. It is assumed that the cold clusters are formed at freeze out. The momentum distribution of a cluster with $M$ nucleons and $Z$ protons for a system with $A$ nucleons is given by
\begin{eqnarray}
\frac{dN_{M}}{d^{3}P}=&& G_{M}{A \choose M} {M \choose Z}\frac{1}{A^{M}}\int \prod_{i=1}^{Z}f_{p}(\textbf{r}_{i},\textbf{p}_{i}) \prod_{i=Z+1}^{M}f_{n}(\textbf{r}_{i},\textbf{p}_{i})                  \nonumber \\
&& \rho^{W}(\textbf{r}_{k_{1}},\textbf{p}_{k_{1}},...,\textbf{r}_{k_{M-1}},\textbf{p}_{k_{M-1}})             \nonumber \\
&& \delta(\textbf{P}-(\textbf{p}_{1}+...+\textbf{p}_{M}))d\textbf{r}_{1}d\textbf{p}_{1}...d\textbf{r}_{M}d\textbf{p}_{M}.
\end{eqnarray}
Here the $f_{n}$ and $f_{p}$ are the neutron and proton phase-space density, which are obtained by performing Wigner transformation based on Gaussian wave packet. The relative coordinate $\textbf{r}_{k_{1}}, ..., \textbf{r}_{k_{M-1}}$ and momentum $\textbf{p}_{k_{1}}, ..., \textbf{p}_{k_{M-1}}$ in the $M-$nucleon rest frame are used for calculating the Wigner density $\rho^{W}$ \cite{Ma97,Ch03}. The spin-isospin statistical factor $G_{M}$ is 3/8, 1/12 and 1/96 corresponding to M=2, 3 and 4, respectively. The root-mean-square radii of intending a cluster is needed for the Wigner density, i.e., 1.61 fm and 1.74 fm for triton and $^{3}$He, loosely bound for hypernuclide.

\section{Results and discussion}

The cluster production in heavy-ion collisions is associated with the baryon ingredient in phase space and with the structure effect. The multiplicity is contributed from the dynamical propagation of the baryons and also from the nuclear fragmentation. The statistical decay of primary fragments dominates the low-energy cluster formation. The nuclear fragmentation reactions have been extensively investigated both in experiments and in theories, in particular on the issues of spinodal multifragmentation, liquid-gas phase transition, properties of highly excited nuclei, symmetry energy at subsaturation densities etc \cite{Ch04,Co93,Po95,Wu98,Ma99}. The nuclear clusters with the charge number Z$\leq$2 manifest the baryonic matter properties in high-energy heavy-ion collisions and the bound state in the hadronization of quark-gluon plasma (QGP), might be the probes of the first order phase transition. On the other hand, the cluster spectra in phase space may shed light on the nuclear equation of state at high-baryon density. Shown in Fig. 1 is the rapidity and kinetic energy spectra of deuteron, triton, $^{3}$He and $^{4}$He in collisions of $^{197}$Au+$^{197}$Au at the incident energy of 1.5 GeV/nucleon. The clusters are mainly produced in the projectile and target-like region, which are contributed from the fragmentation reaction in semicentral collisions. It is noticed that the particles ($\pi$, $\eta$, K, $\Lambda$, $\Sigma$ etc) are created in the midrapidity domain. The peak in the energy spectra is caused from the fermi motion and hadron-hadron collisions.

\begin{figure}
\begin{center}
{\includegraphics*[width=0.95\textwidth]{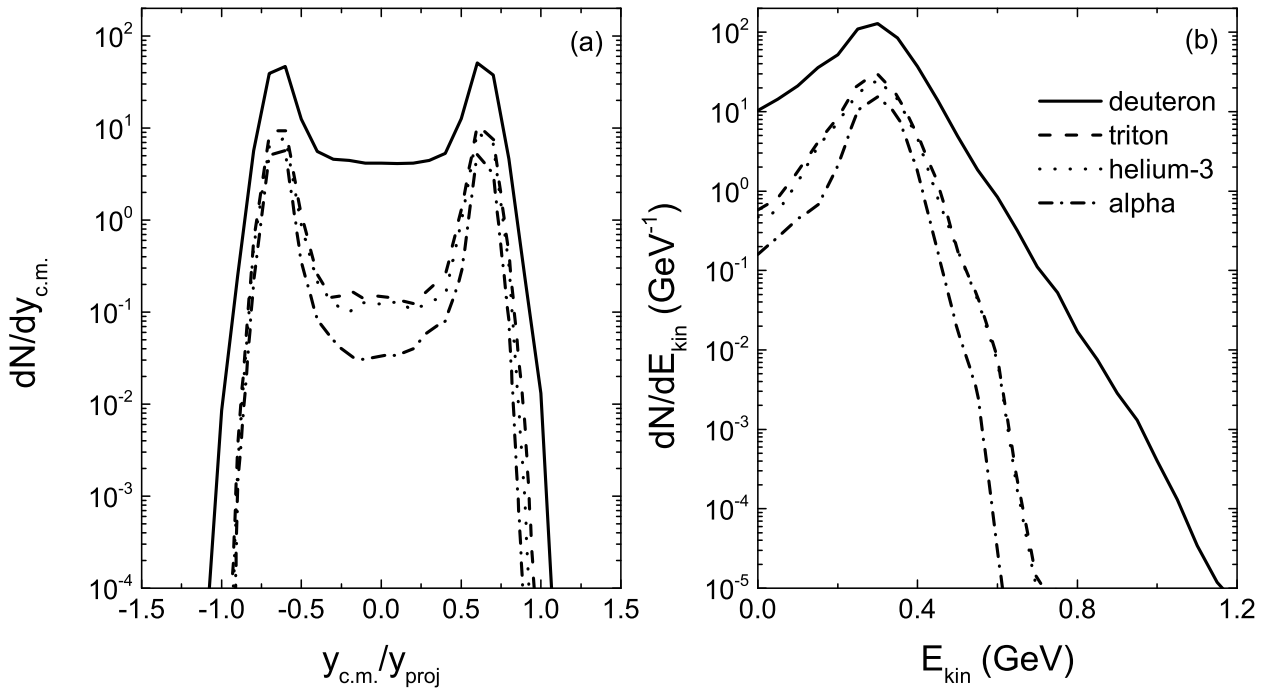}}
\end{center}
\caption{(a) Rapidity and (b) kinetic energy distributions of deuteron, triton, $^{3}$He and $\alpha$ in collisions of $^{197}$Au+$^{197}$Au at the incident energy of 1.5 GeV/nucleon.}
\end{figure}

\begin{figure}
\begin{center}
{\includegraphics*[width=0.95\textwidth]{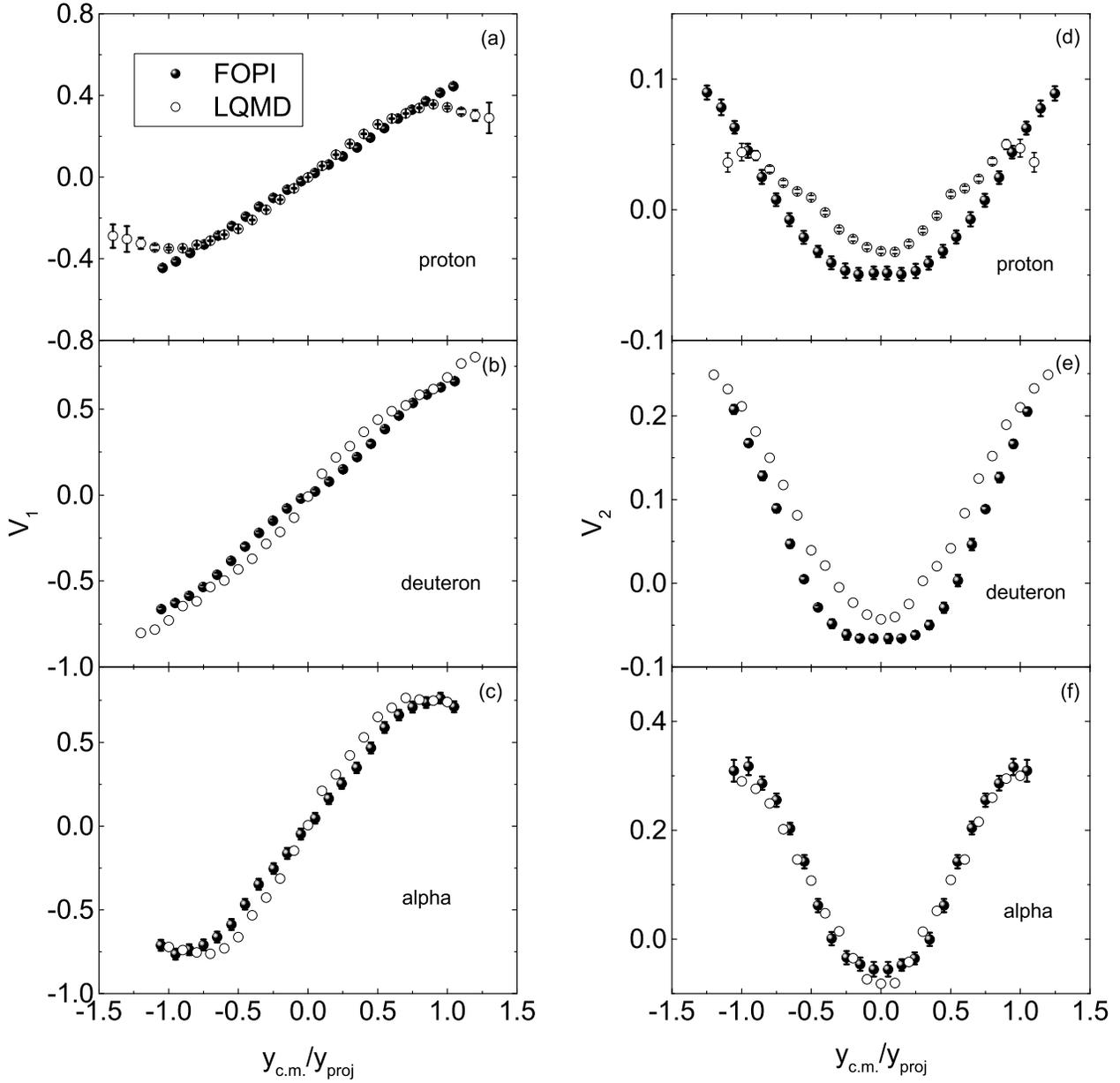}}
\end{center}
\caption{Directed and elliptic flows of proton, deuteron and alpha as functions of rapidities in collisions of $^{197}$Au+$^{197}$Au within at centrality of 3.35 fm $<b<$ 6 fm and incident energy of 1.5 GeV/nucleon. The experimental data from FOPI collaboration is shown for comparison \cite{Re12}.}
\end{figure}

The collective flows are nice probes for extracting the properties of high-density matter formed in heavy-ion collisions, in-medium hadrons, hadron-hadron scattering, decay of resonances and reabsorption of meson etc. The flow information can be expressed as the first and second coefficients from the Fourier expansion of the azimuthal distribution with angle $\phi$ as $\frac{dN}{d\phi}\left(y,p_t\right) = N_0 \left[1 + 2V_1\left(y,p_t\right)cos\left(\phi\right)+2V_2\left(y,p_t\right)cos\left(2\phi\right)\right]$, where $
p_t=\sqrt{p_{x}^{2}+p_{y}^{2}}$ and $y$ are the transverse momentum and the longitudinal rapidity along the beam direction, respectively. The directed flow and the elliptic flow respectively are $V_1=<p_x/p_t>$ and $V_2 = <(p_{x}^{2}-p_{y}^{2})/p_{t}^{2}>$, so the directed flow provides information on the azimuthal anisotropy of the transverse emission and the elliptic flow gives the competition between the in-plane ($V_2 > 0$) and out-of-plane ($V_2 < 0$, squeeze-out) emissions. As a test of the model, we calculated the rapidity distributions of directed and elliptic flows of protons, deuterons and $\alpha$ in the $^{197}$Au + $^{197}$Au collisions and compared with the experimental data from the FOPI collaboration \cite{Re12} as shown in Fig. 2. The calculated results are nicely consistent with the available data at GSI. The slope of the transverse flow increases with the mass number of cluster, which is caused from the contribution of spectator nucleons on the cluster formation. The U-shape elliptic flows become more steep for the $\alpha$ emission. The out-of-plane emissions in the midrapidity region are pronounced for the cluster formation, which are created during the compression state in heavy-ion collisions.

\begin{figure}
\begin{center}
{\includegraphics*[width=0.95\textwidth]{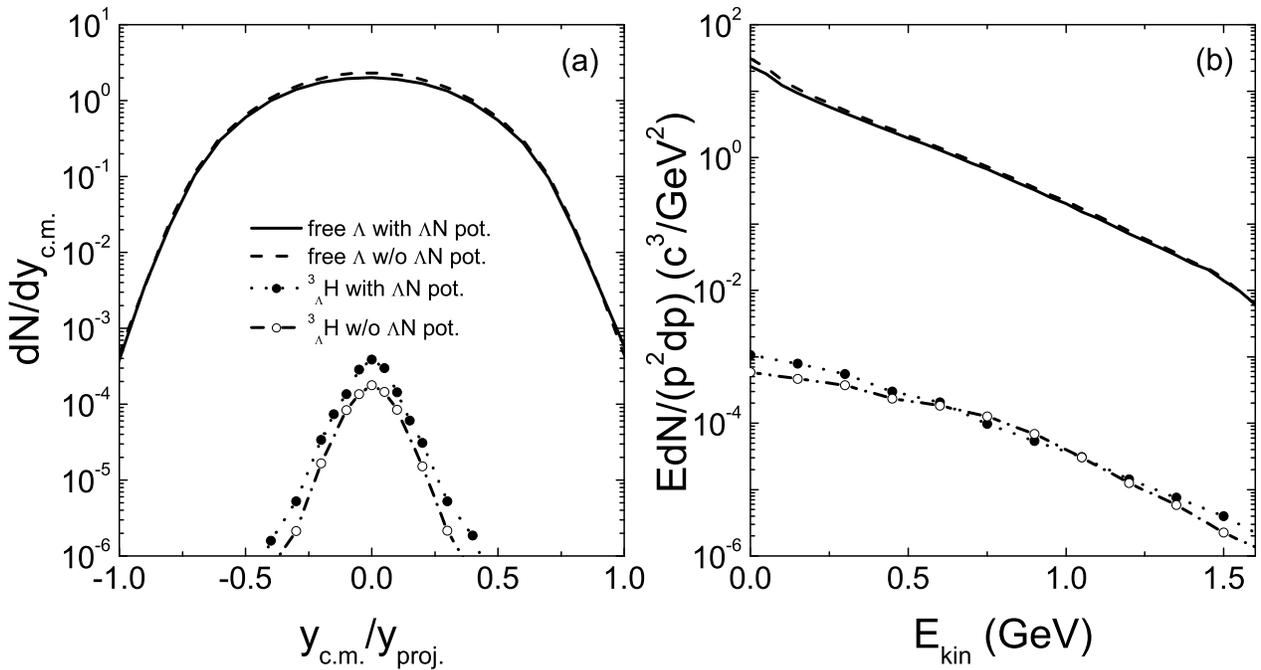}}
\end{center}
\caption{Rapidity and invariant spectra of free $\Lambda$ and light hypernuclide $^{3}_{\Lambda}$H in the central collisions of $^{197}$Au+$^{197}$Au at the incident energy of 2 GeV/nucleon.}
\end{figure}

The nuclear chart is extended to the three dimensional structure by implementing strangeness degree of freedom, which exhibits some interesting phenomena, i.e., the neutral bound states (nn$\Lambda$, n$\Lambda\Lambda$), new spectroscopy etc. The heavy-ion collisions provide a unique way to produce the neutron-rich or proton-rich hypernuclides in the terrestrial laboratories. Shown in Fig. 3 is a comparison of free $\Lambda$ and hypernuclide $^{3}_{\Lambda}$H produced in the central collisions of $^{197}$Au + $^{197}$Au at the incident energy of 2 GeV/nucleon. The hyperfragments are formed when the relative distance between a $\Lambda$ and nucleon is enough to form the bound state and the yields are strongly suppressed in comparison with free $\Lambda$. The attractive $\Lambda-$nucleon potential is favorable for the hypernuclide formation, but has negligible contribution for free $\Lambda$ production. The light hypernuclides $^{3}_{\Lambda}$H and $^{4}_{\Lambda}$H in the reaction $^{6}$Li+$^{12}$C at an incident energy of 2 GeV/nucleon were measured by the HypHI collaboration \cite{Ra15}.

\begin{figure}
\begin{center}
{\includegraphics*[width=0.95\textwidth]{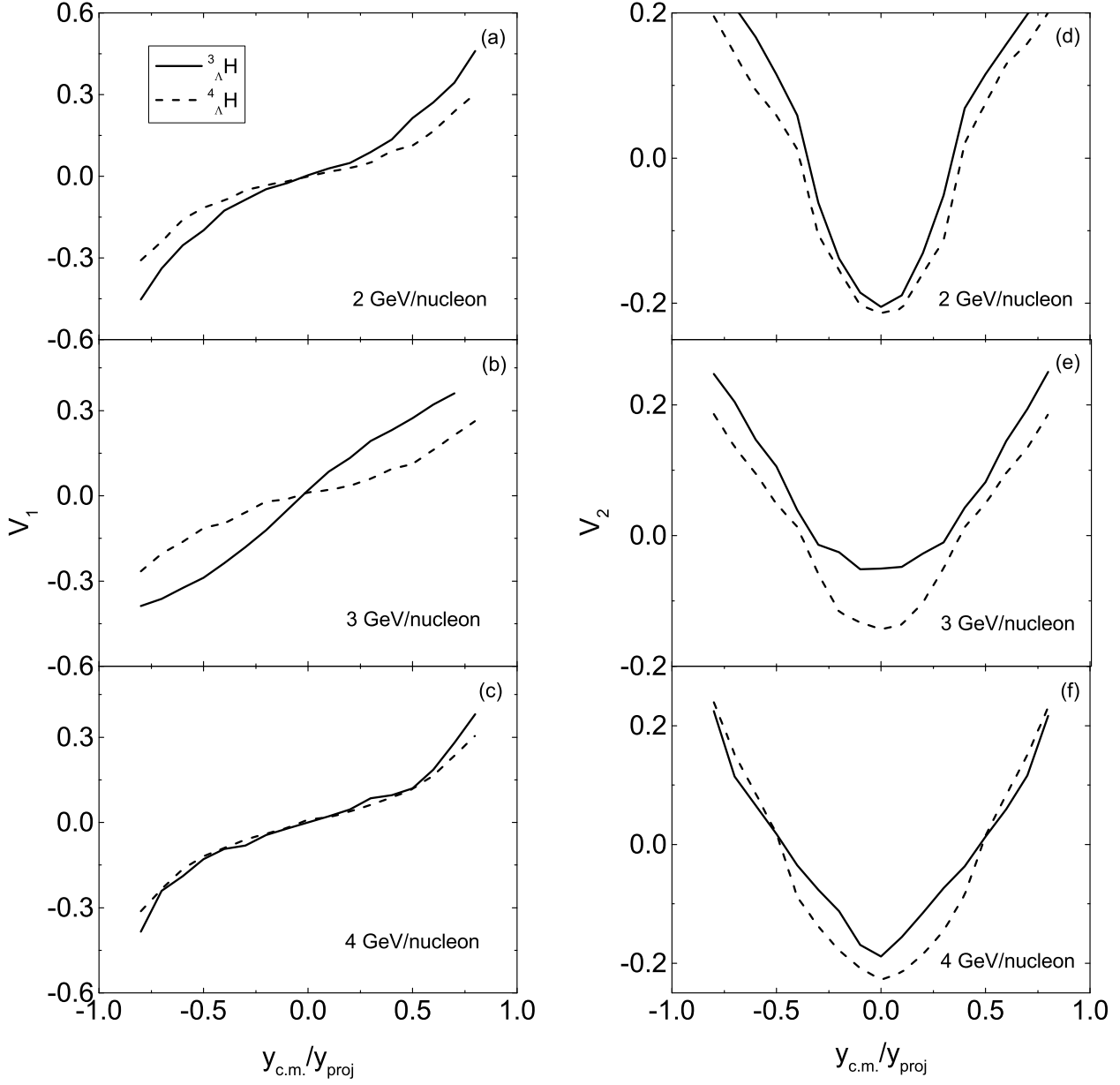}}
\end{center}
\caption{Rapidity distributions of directed and elliptic flows for $^{3}_{\Lambda}$H and $^{4}_{\Lambda}$H produced in collisions of $^{197}$Au+$^{197}$Au at 2, 3 and 4 GeV/nucleon, respectively.}
\end{figure}

The hypernuclear physics is one of the topical issues in the future HIAF facility, in particular the exotic hypernuclei in extremely neutron-rich or proton-rich region, multiple strangeness nuclei, neutral hyperclusters ($_{\Lambda}^{3}$n, $_{\Lambda\Lambda}^{4}$n) and high-density symmetry energy from the isospin ratios of strange particles etc. The azimuthal emission of hypernuclide produced in heavy-ion collisions manifests the structure effect and high-baryon matter properties. The short-range correlation of hyperon-nucleon interaction and multiple collisions contribute the hypernuclide production. On the other hand, the density dependent symmetry energy influences the isospin diffusion in heavy-ion collisions, in particular in the high-density domain, which consequently dominates the neutron-rich hypernucleus formation. Shown in Fig. 4 is a comparison of collective flows of $^{3}_{\Lambda}$H and $^{4}_{\Lambda}$H in collisions of $^{197}$Au+$^{197}$Au at the incident energy of 2, 3 and 4 GeV/nucleon, respectively. The similar structure to nuclear clusters is found, which manifests the formation of hyperfragments following the cluster mechanism in the fragmentation process. It is noticed that the hyperfragments are mainly produced via the capture of free hyperon by spectator nucleons. The direct 3-body or 4-body collisions associated with hyperon and nucleons are not implemented in the calculation, which enable the midrapidity and energetic hyperfragment production. On the other hand, it should be mentioned that the hyperfragments formed in the preequilibrium process are not taken into account in the LQMD model, which have been manifested the significant contribution on the hyperfragment formation from the parton-hadron quantum molecular dynamics (PHQMD) calculations \cite{Ai20}. Modifications of the both shortcomings are in progress.

\begin{figure}
\begin{center}
{\includegraphics*[width=0.95\textwidth]{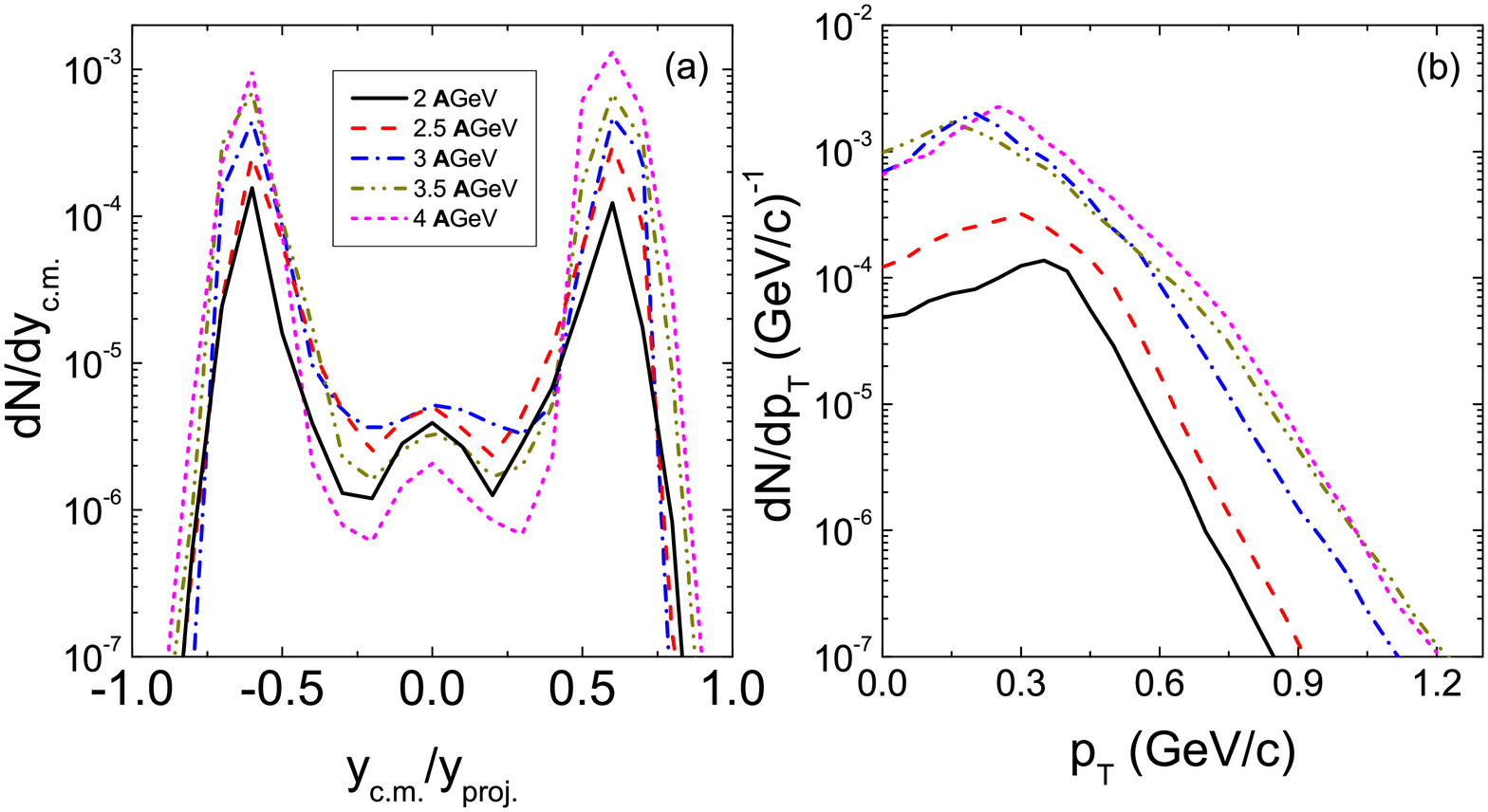}}
\end{center}
\caption{Comparison of rapidity and transverse momentum spectra of $^{3}_{\Lambda}$H at different energies.}
\end{figure}

The kinematics of hyperfragments produced in heavy-ion collisions is helpful in managing the detector system in the experiments. The rapidity and transverse momentum spectra of hypercluster $^{3}_{\Lambda}H$ are calculated as shown in Fig. 5. It is obvious that the light hyperfragments are formed in the projectile or target-like region at different energies. All collision centralities within the impact parameter $b_{max}=R_{0}(A_{P}^{1/3}+A_{T}^{1/3})$ with $R_{0}=0.16$ fm have been included in the calculation. It is indicated that the hyperon is captured by spectator nucleons to form the hyperfragments and the velocity is reduced in comparison with the colliding partners. The transverse momentum spectra become broader with increasing the incident energy because of the energetic hyperons. Once the beam energy above 3 GeV/nucleon, the hyperfragment production is close to the saturation yield, which is favorable for measuring the hypernuclides at HIAF energies. It has advantage that the extremely neutron-rich hypernuclei might be created via heavy-ion collisions. The first attempt at HIAF is planned to investigate the light hypernuclei ($^{3}_{\Lambda}$H, $^{4}_{\Lambda}$H, $^{4}_{\Lambda}$He, $^{4}_{\Lambda\Lambda}$H etc) in collisions of $^{20}$Ne+$^{12}$C at the beam energy of 4.25 GeV/nucleon, which are economical for detecting in experiments because of the large Lorentz factor. It is calculated from the LQMD model that the hyperclusters are formed in the projectile- and target-like rapidity regime with the yields of several tens $\mu$b \cite{Fe20b}.

\section{Conclusions}

In summary, the emission mechanism of nuclear clusters and light hypernuclei in heavy-ion collisions is investigated within the LQMD transport model. The Wigner density approach is used for recognizing the clusters at the stage of freeze out in the nuclear reactions. The elliptic flows of nuclear clusters from FOPI collaboration are well reproduced with the approach. The nuclear clusters and hyperfragments are formed in the projectile and target-like region. Although the influence of hyperon-nucleon potential on the free hyperons is negligible, it is favorable for the bound hyperfragment formation. The directed and elliptic flows of hypernuclides $^{3}_{\Lambda}$H and $^{4}_{\Lambda}$H manifest the similar structure with the nuclear clusters. The energy dependence of hyperfragment production is weak above 3 GeV/nucleon and a broad momentum spectrum is pronounced for the light hypernuclei. The collective flows of light hypernuclei in heavy-ion collisions are associated with the properties of hadronic matter, hypernuclear structure, hyperon-nucleon interaction and multiple collisions. Future experiments at the next-generation high-intensity accelerator are to investigate the exotic hypernuclear physics.

\textbf{Acknowledgements}
This work was supported by the National Natural Science Foundation of China (Projects No. 11722546 and No. 11675226) and the Talent Program of South China University of Technology.

\end{document}